\documentclass[reprint, showpacs, superscriptaddress, prl, aps, longbibliography]{revtex4-1}

\usepackage{graphicx}
\usepackage{dcolumn}
\usepackage{bm}
\usepackage{multirow}
\usepackage{siunitx}
\DeclareSIUnit\torr{Torr}
\usepackage{amsmath,amssymb}
\usepackage[version=4]{mhchem}

\begin{document}
	
	\preprint{APS/123-QED}
	\title{Damping's effect on the magnetodynamics of spin Hall nano-oscillators}

	\author{Yuli Yin}
	\email{yuri@seu.edu.cn}
	\affiliation{Department of Physics, Southeast University, 211189 Nanjing, China}
	\affiliation{Department of Physics, University of Gothenburg, 412 96 Gothenburg, Sweden}
	
	\author{Philipp D\"{u}rrenfeld}
	\affiliation{Department of Physics, University of Gothenburg, 412 96 Gothenburg, Sweden}
	
	\author{Mykola Dvornik}
	\affiliation{Department of Physics, University of Gothenburg, 412 96 Gothenburg, Sweden}
	
	\author{Martina Ahlberg}
	\affiliation{Department of Physics, University of Gothenburg, 412 96 Gothenburg, Sweden}
	
	\author{Afshin Houshang}
	\affiliation{Department of Physics, University of Gothenburg, 412 96 Gothenburg, Sweden}
	
	\author{Ya Zhai}
	\affiliation{Department of Physics, Southeast University, 211189 Nanjing, China}
	
	\author{Johan {\AA}kerman}
	\affiliation{Department of Physics, University of Gothenburg, 412 96 Gothenburg, Sweden}
	\affiliation{Department of Materials and Nano Physics, School of Information and Communication Technology, KTH Royal Institute of Technology, Electrum 229, 164 40 Kista, Sweden}

	\date{\today}
	
	\begin{abstract}  
		We study the impact of spin wave damping ($\alpha$) on the auto-oscillation properties of nano-constriction based spin Hall nano-oscillators (SHNOs). The SHNOs are based on a 5\,nm Pt layer interfaced to a 5\,nm Py$_{100-x-y}$Pt$_{x}$Ag$_{y}$ magnetic layer, where the Pt and Ag contents are co-varied to keep the saturation magnetization constant (within \SI{10}{\percent}), while $\alpha$ varies close to a factor of three. We systematically investigate the influence of the Gilbert damping on the magnetodynamics of these SHNOs by means of electrical microwave measurements. Under the condition of a constant field, the threshold current scales with the damping in the magnetic layer. The threshold current as a function of field shows a parabolic-like behavior, which we attribute to the evolution of the spatial profile of the auto-oscillation mode. The signal linewidth is smaller for the high-damping materials in low magnetic fields, although the lowest observed linewidth was measured for the alloy with least damping.
	\end{abstract}
	
	\maketitle
	
	\section{\label{sec:level1}Introduction}
	
	Spin Hall nano-oscillators (SHNO) are spintronic devices in which magnetization oscillations are induced by pure spin currents~\cite{Chen2016procieee}. These pure spin currents can be experimentally realized via the spin Hall effect (SHE) in an adjacent heavy metal layer~\cite{liu2012sc,demidov2012ntm,demidov2017pr} or by non-local spin injection~\cite{demidov2015apl,Haidar2016prb}. SHNOs, which use the SHE in a heavy metal layer, have been fabricated in a variety of device layouts, which all utilize the focusing of charge current into a region with a lateral size of tens to hundreds of nanometers. This focusing is commonly done via a nano-gap between two highly conductive electrodes~\cite{demidov2012ntm,liu2013prl,ranjbar2014ieeeml}, with a nanoconstriction~\cite{divinskiy2017apl,demidov2014apl,duerrenfeld2017ns,mazraati2016apl,dvornik2018}, or with a nanowire~\cite{duan2014ntc,yang2015srp}.
	Most recently, nanoconstriction-SHNOs have attracted large interest, due to their relative ease of fabrication, their direct optical access to the magnetization oscillation area, and their potential for large scale and large distance synchronization of multiple SHNOs~\cite{kendziorczyk2016prb,awad2017ntp}.
	
	Nanoconstriction-SHNOs consist of a bilayer of a ferromagnetic free layer and a SHE inducing heavy metal layer. Since the SHE and the concomitant spin accumulation at the bilayer interface are only influenced by the current density in the heavy metal layer, magnetization oscillations of the device under a constant current can be directly linked to the magnetodynamic properties of the magnetic free layer. Until now, the variety of materials from which SHNOs has been fabricated is limited to a few standards like permalloy (Py, \ce{Ni80Fe20}), (Co,Fe)B, or yttrium iron garnet (YIG). However, these materials are different from each other in every one of the key magnetodynamic parameters, such as magnetization~($M$), Gilbert damping~($\alpha$), or exchange constant~($A$).
	
	In a recent study, we have shown how the magnetodynamic properties of Py can be engineered by alloying with the noble metals Pt, Au, and Ag~\cite{yin2015prb}. While alloying with Pt leads to a large increase in damping but only a small decrease in magnetization, alloying with Ag has only a weak effect on the damping but reduces the magnetization relatively strongly. Co-alloying with both elements Pt and Ag thus results in Py$_{100-x-y}$Pt$_{x}$Ag$_{y}$ films, whose $M$ and $\alpha$ can be tuned independently, e.g. the magnetization can be kept constant, while the damping is strongly increased with increasing Pt concentration.
	
	Here, we employ a series of alloyed Py$_{100-x-y}$Pt$_{x}$Ag$_{y}$ thin films in nanoconstriction-SHNOs, where we vary the effective damping of the free layer by a factor of three, while we keep the magnetization of the films constant. Based on these films, we fabricate geometrically identical nanoconstriction-SHNOs and compare their microwave auto-oscillation characteristics. This allows us to directly analyze the influence of one single magnetodynamic property, namely the Gilbert damping, on the spectral characteristics, i.e. the onset current ($I_{\text{DC}}^{\text{th}}$), the output power ($P$), and the linewidth ($\Delta f$).
	
	\section{\label{sec:level2}Spin Hall Nano-Oscillator Devices}
	
	\begin{figure}[t]
		\centering
		\includegraphics[width=3.3in]{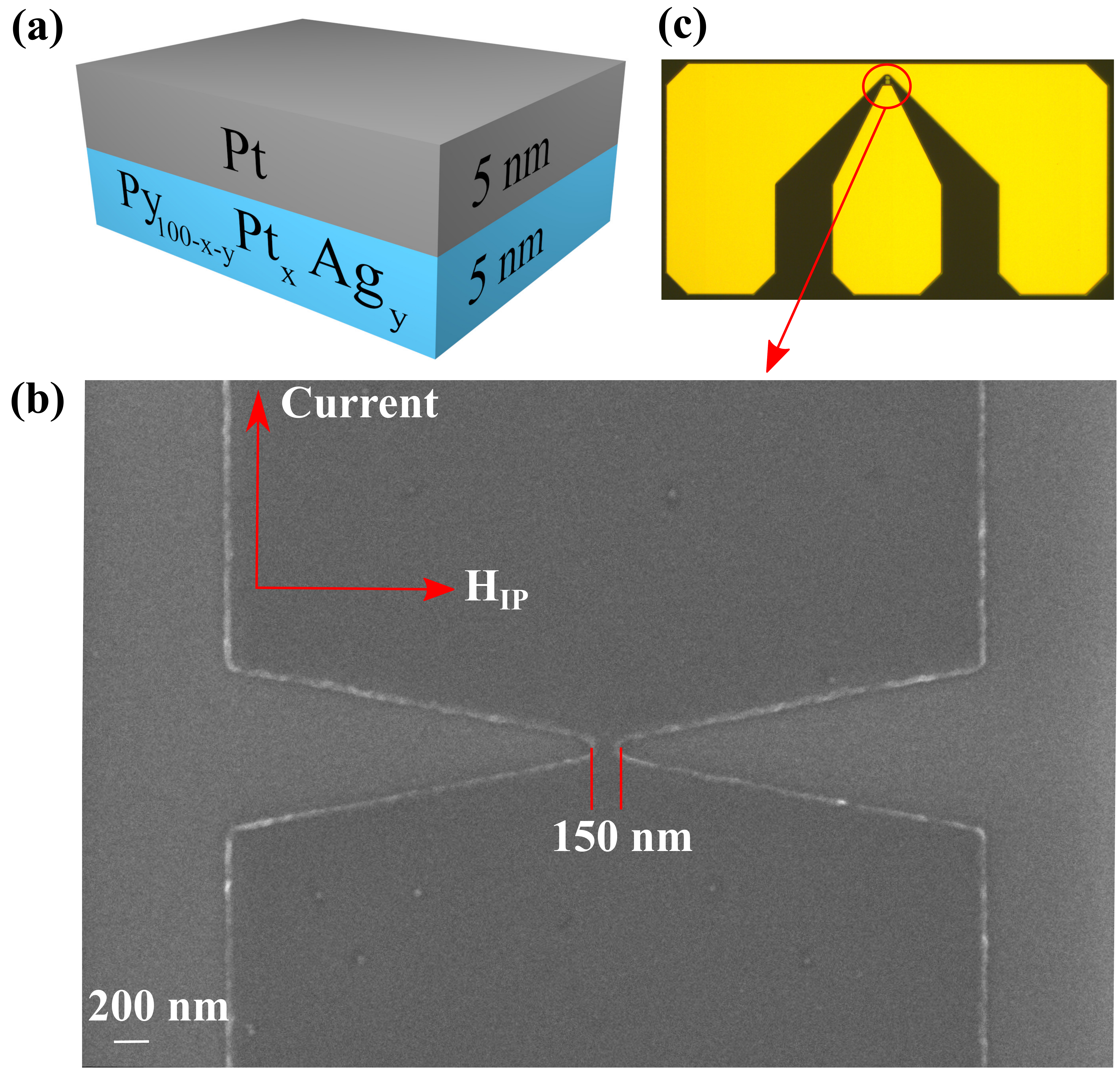}
		\caption{\label{fig:fig1} (a) Schematic representation of the sputtered bilayer structure. (b) SEM micrograph of a nanoconstriction-SHNO showing the relative orientations of current and field. (c) Optical micrograph showing the microwave wave guide used for contacting the SHNOs.}
	\end{figure}
	
	Bilayers of \SI{5}{\nm} Py$_{100-x-y}$Pt$_{x}$Ag$_{y}$ and \SI{5}{\nm} Pt were sputter-deposited onto sapphire substrates in a high-vacuum chamber with a base pressure of less than \SI{3e-8}{\torr}. The deposition was carried out with \SI{3}{\milli\torr} argon gas at a flow rate of 30\,sccm. The alloyed layers were co-sputtered from up to 3 targets, and the Py target power was kept constant at \SI{350}{\W}, while the noble metal sputtering powers and the sputtering time was adjusted for composition and thickness, respectively. The top Pt layer was magnetron sputtered with a dc power of \SI{50}{\W}.
	The alloy compositions are Py$_{84}$Ag$_{16}$ (S01), Py$_{77.5}$Pt$_{10}$Ag$_{12.5}$ (S02), Py$_{75}$Pt$_{15}$Ag$_{10}$ (S03), and Py$_{73}$Pt$_{19}$Ag$_{8}$ (S04), chosen to result in a constant saturation magnetization throughout the series of SHNOs~\cite{yin2015prb}.
	
	Devices for electrical measurements were fabricated from these bilayers by electron beam lithography and argon ion beam etching, using the negative resist as an etching mask. Nanoconstrictions were formed by two symmetrical indentations with a \SI{50}{\nm} tip radius into \SI{4}{\um} wide stripes, see Fig.~\ref{fig:fig1}(b). The width of the nanoconstrictions is \SI{150}{\nm}. Finally, \SI{1}{\um} thick copper waveguides with a \SI{150}{\um} pitch were fabricated by optical lithography and lift-off, see Fig.~\ref{fig:fig1}(c).
	
	\section{\label{sec:level3}Film Characterization}
	
\begin{figure}[t]
		\centering
		\includegraphics[width=3.3in]{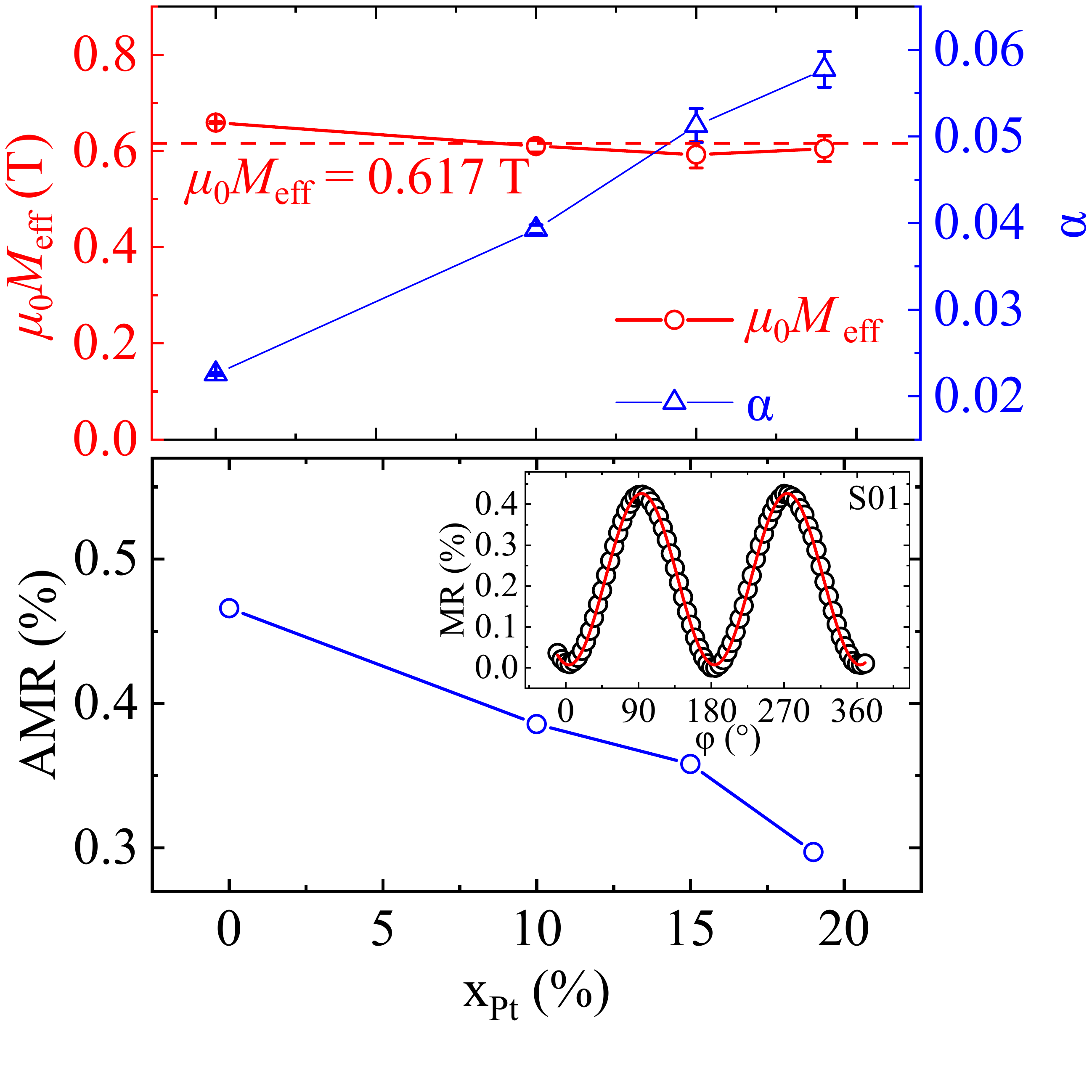}
		\caption{\label{fig:fig2} (a) Magnetization and damping of the alloyed films in the bilayer as measured by CPW-based FMR. (b) AMR of the extended layer structure. The inset shows the angular-dependent relative resistance of the Py$_{84}$Ag$_{16}$/Pt (S01) bilayer, together with a fit to a $\cos^2$-function.}
	\end{figure}

	Characterization of the extended bilayer samples was performed by ferromagnetic resonance (FMR), and two-point anisotropic magneotresistance (AMR) measurements.
	The FMR was carried out with in-plane applied fields using a NanOsc Instruments PhaseFMR with a \SI{200}{\um} wide coplanar waveguide (CPW). An asymmetric Lorentzian was fit to the absorption peaks. The frequency dependence of the determined resonance fields and linewidths was subsequently used to extract the effective magnetization ($\mu_{0}M_{\text{eff}}$) and the damping parameter ($\alpha$), respectively~\cite{yin2015prb}.  Figure~\ref{fig:fig2}(a) shows the two parameters, $\mu_{0}M_{\text{eff}}$ and $\alpha$, as a function of Pt concentration. The magnetization is constant throughout the sample series ($\mu_{0}M_{\text{eff}}$~=~0.617(34)\,T), while the damping increases linearly from  $0.023(1)$ to $0.058(3)$ as the Pt concentration increases from 0 (Py$_{84}$Ag$_{16}$) to \SI{19}{\percent} (Py$_{73}$Pt$_{19}$Ag$_{8}$). The small layer thickness compared to the films in Ref.~\onlinecite{yin2015prb} results in a slightly lower magnetization, whereas the damping is enhanced as a consequence of spin pumping into the adjacent Pt layer~\cite{tserkovnyak2002prl,mizukami2002prb,sun2013prl}.
	
	The AMR was determined by probing the resistance of \SI{4}{\um} wide stripes in a rotating \SI{90}{\milli\tesla} in-plane magnetic field. A representative AMR measurement is presented in the inset of Fig.~\ref{fig:fig2}(b), together with a fit of a $\cos^2$-function to the data. The angle $\varphi=\ang{0}$ denotes a perpendicular orientation between current and field, and the AMR (Fig.~\ref{fig:fig2}(b)) is calculated by the difference in resistance at perpendicular and parallel alignments via $\text{AMR}=\frac{R_{\parallel}-R_{\perp}}{R_{\perp}}$. The AMR is below \SI{1}{\percent}, which is a result of the majority of the current flowing through the nonmagnetic platinum layer, which has a higher conductivity than the Py alloys. The AMR reduces by $\approx \SI{30}{\percent}$ across the samples series, but the absolute resistance of the bilayers decreases by less than \SI{5}{\percent}. The AMR magnitude is therefore most likely governed by the alloy composition, since the amount of current in the magnetic layer does not change significantly.
	
	\section{\label{sec:level4}Microwave Emission Measurements and Discussion}
	
	\begin{figure}[t]
		\centering
		\includegraphics[width=3.3in]{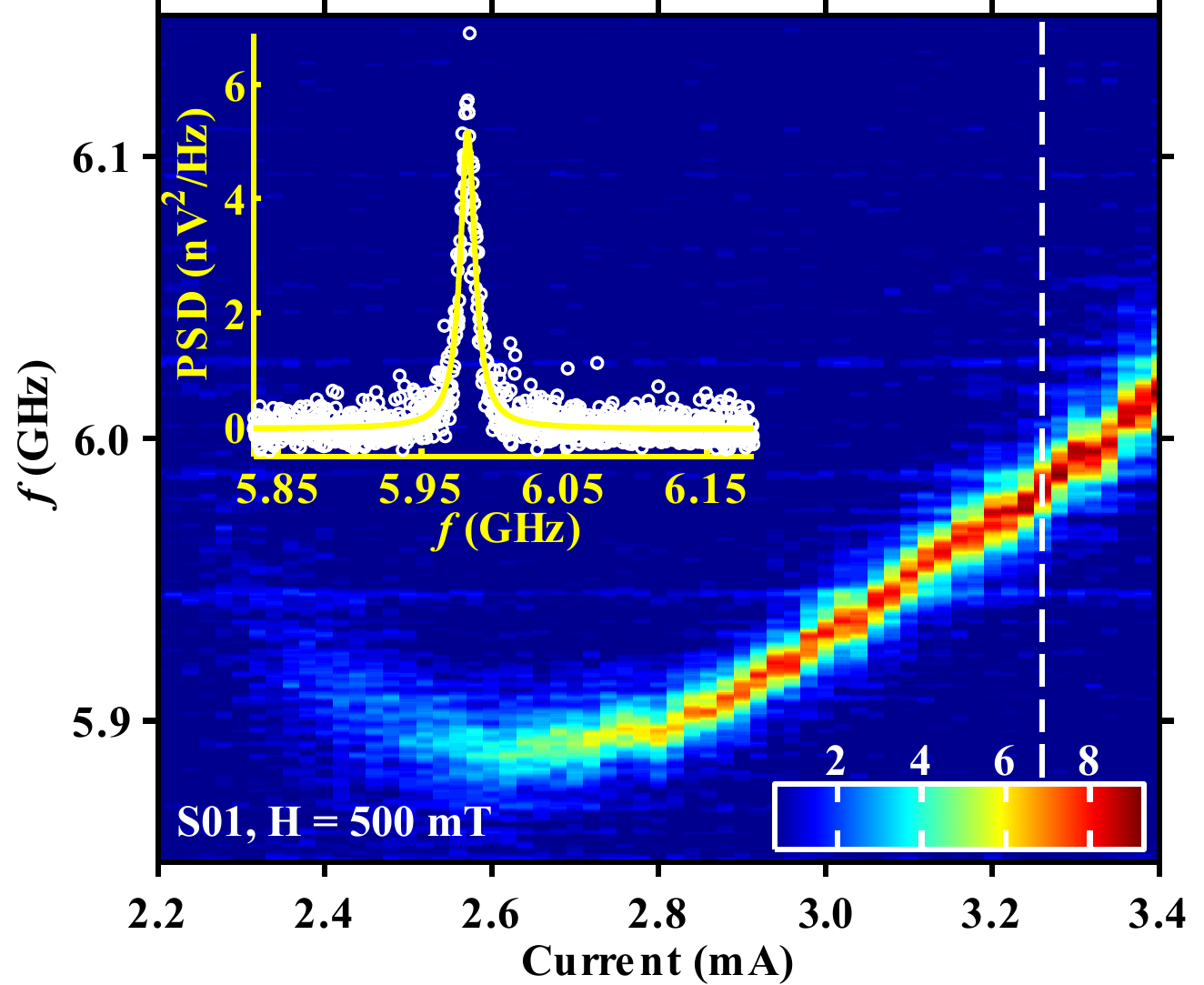}
		\caption{\label{fig:fig3} Power spectral density (PSD) of the Py$_{84}$Ag$_{16}$/Pt (S01) SHNO as a function of current in an external field of $\mu_{0}H_{\text{ext}} = \SI{0.5}{\tesla}$, tilted \ang{80} OOP. The inset shows the PSD at $I_{\text{DC}} = \SI{3.26}{\mA}$ and the solid line is a Lorentzian fit resulting in $\Delta f = \SI{5.98}{\MHz}$ and $P = \SI{1.02}{\pico\watt}$.}
	\end{figure}
	
	The microwave measurements were conducted with the devices placed in a magnetic field oriented at an out-of-plane (OOP) angle of \ang{80} from the film plane, and an in-plane angle of $\varphi = \ang{0}$. The in-plane component of the magnetic field ($H_{\text{ext}}^{\text{IP}}$) was thus perpendicular to the current flow direction, as sketched in Fig.~\ref{fig:fig1}(b). The relative orientation of the current and $H_{\text{ext}}^{\text{IP}}$ yields a spin-torque caused by the spin current from the Pt layer, which reduces the damping in the Py layer and leads to auto-oscillations for sufficiently large positive applied dc currents ($I_{\text{DC}}$)~\cite{liu2011prl}. The current was applied to the samples via the dc port of a bias-tee and the resulting microwave signals from the devices were extracted from the rf port of the bias-tee. The microwave signals were then amplified by a broadband (\num{0.1} to \SI{40}{\GHz}) low-noise amplifier with a gain of +32\,dB before being recorded by a spectrum analyzer (Rohde\&Schwarz FSV-40) with a resolution bandwidth of \SI{500}{\kHz}. All measurements were carried out at room temperature.
	
	A typical microwave measurement of a Py$_{84}$Ag$_{16}$/Pt (S01) device in a constant field of $\mu_{0}H_{\text{ext}} = \SI{0.5}{\tesla}$ and a varying current is displayed in Fig.~\ref{fig:fig3}. The peak frequency first decreases slightly after the oscillation onset at $I_{\text{DC}}^{\text{th}} = \SI{2.26}{\mA}$, then reaches a minimum at $\sim \SI{2.6}{\mA}$, and finally increases up to the maximum applied current of \SI{3.4}{\mA}. A Lorentzian peak function fits well to the auto-oscillation signal, see inset of Fig.~\ref{fig:fig3}, allowing for determination of the full-width at half-maximum linewidth ($\Delta f$) and the integrated output power ($P$). Besides the highly coherent auto-oscillation mode, no other modes are excited under these field conditions.
	
	\begin{figure}[t]
		\centering
		\includegraphics[width=3.3in]{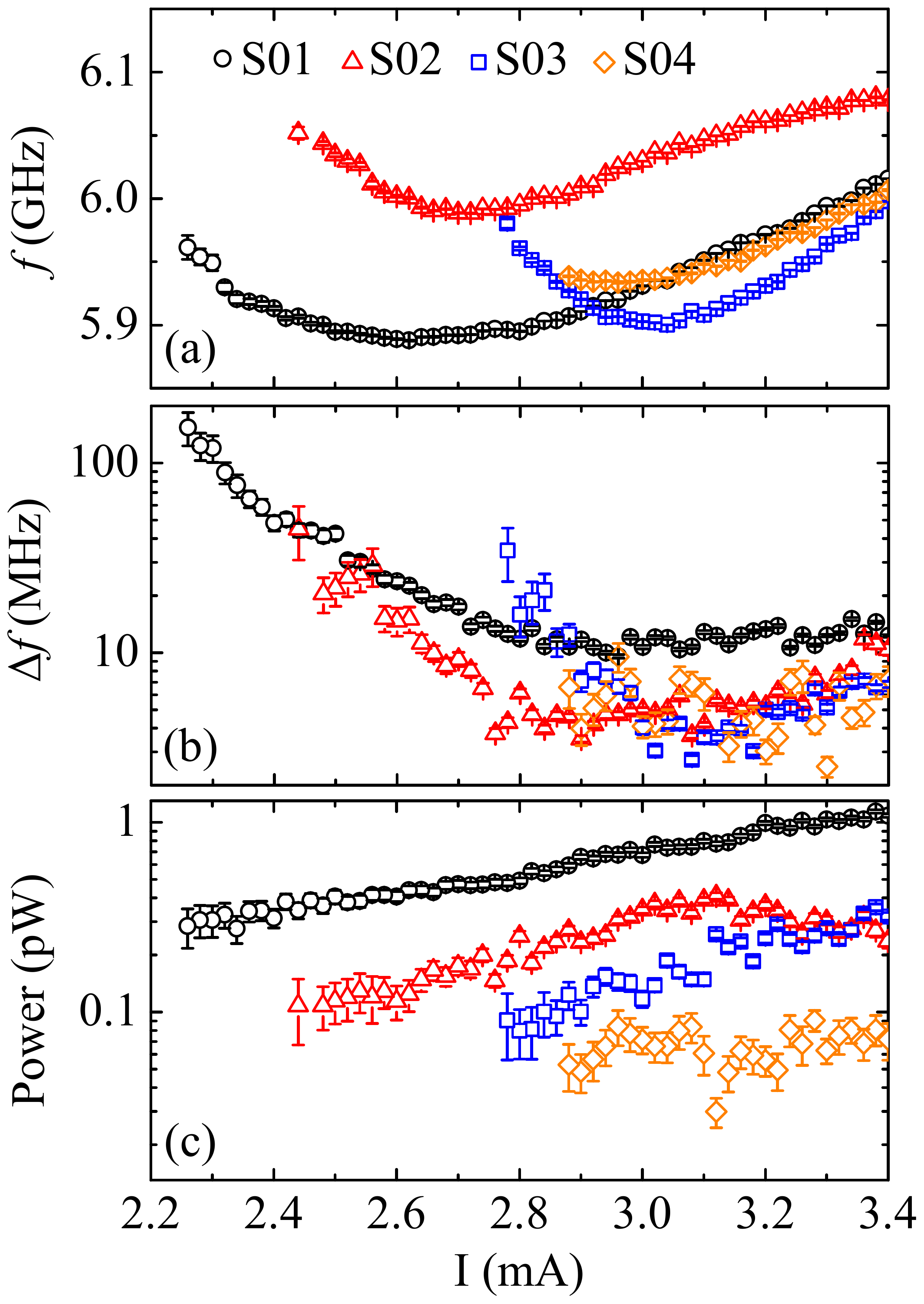}
		\caption{\label{fig:fig4} (a) Frequency, (b) linewidth, and (c) integrated power of the microwave auto-oscillations as a function of current for four different SHNOs with increasing damping. The applied field is $\mu_{0}H_{\text{ext}} = \SI{0.5}{\tesla}$, tilted \ang{80} OOP.}
	\end{figure}
	
	Figure~\ref{fig:fig4} shows the determined auto-oscillation characteristics of SHNOs with different alloy composition and damping. The measurements were again made in a constant field of $\SI{0.5}{\tesla}$. The oscillation frequencies in Fig.~\ref{fig:fig4}(a) lie around 6.0$\pm$0.1\,GHz for all samples, and the current-frequency dependence is virtually identical above the individual threshold currents. However, the current range where $f$ decreases with $I_{\text{DC}}$ is missing for the S04 sample, which suggests that the threshold current is underestimated for this device. The comparable frequencies of all samples confirm that the saturation magnetization is constant throughout the alloy series. Furthermore, the quantitatively similar current tunability implies that the increased damping does not change the fundamental nature of the excited auto-oscillation mode.
	
	The linewidth of the SHNOs decreases rapidly after the auto-oscillation onset and then levels off for higher $I_{\text{DC}}$ values, as shown in Fig.~\ref{fig:fig4}(b). This behavior is consistent with previous studies on nanoconstriction-SHNOs made of permalloy films~\cite{awad2017ntp,duerrenfeld2017ns}. The low-damping device S01 reaches its minimum level at $\Delta f \sim \SI{11}{\MHz}$, while the SHNOs with higher damping materials all have a similar minimum linewidth of $\Delta f \sim \SI{5}{\MHz}$. The linewidth is inversely proportional to the mode volume~\cite{slavin2009ieeem}, and the decrease in $\Delta f$ can therefore be attributed to a spatial growth of the auto-oscillation region as the damping increases. Nevertheless, the active area of the device is confined to the nanoconstriction, which limits the reduction in linewidth. 
	
	The output power of the four nanoconstriction-SHNOs is shown as a function of $I_{\text{DC}}$ in Fig.~\ref{fig:fig4}(c). The power grows almost exponentially with increasing current for all samples. However, $P$ drops dramatically as the Pt concentration increases. The AMR also decreases in the higher damping samples, but the reduction is too small to fully account for the drop in power. Together with the trend in linewidth, the evolution of the power contradicts the general assumption $\Delta f \propto \alpha / P$~\cite{slavin2009ieeem, tiberkevich2009apl, tiberkevich2014srp}. This equation is only valid in the vicinity of the threshold current and a direct comparison to the data is problematic,  due to the experimental difficulties of determining $I_{\text{DC}}^{\text{th}}$. Still, the direct relation between the \emph{intrinsic oscillator} power and the \emph{electrically measured} power is put into question due to the remarkable decrease in the measured $P$. A number of factors could influence the signal strength, e.g. rectification, spin-pumping, and the inverse spin-Hall effect.

	\begin{figure}[t]
		\centering
		\includegraphics[width=3.3in]{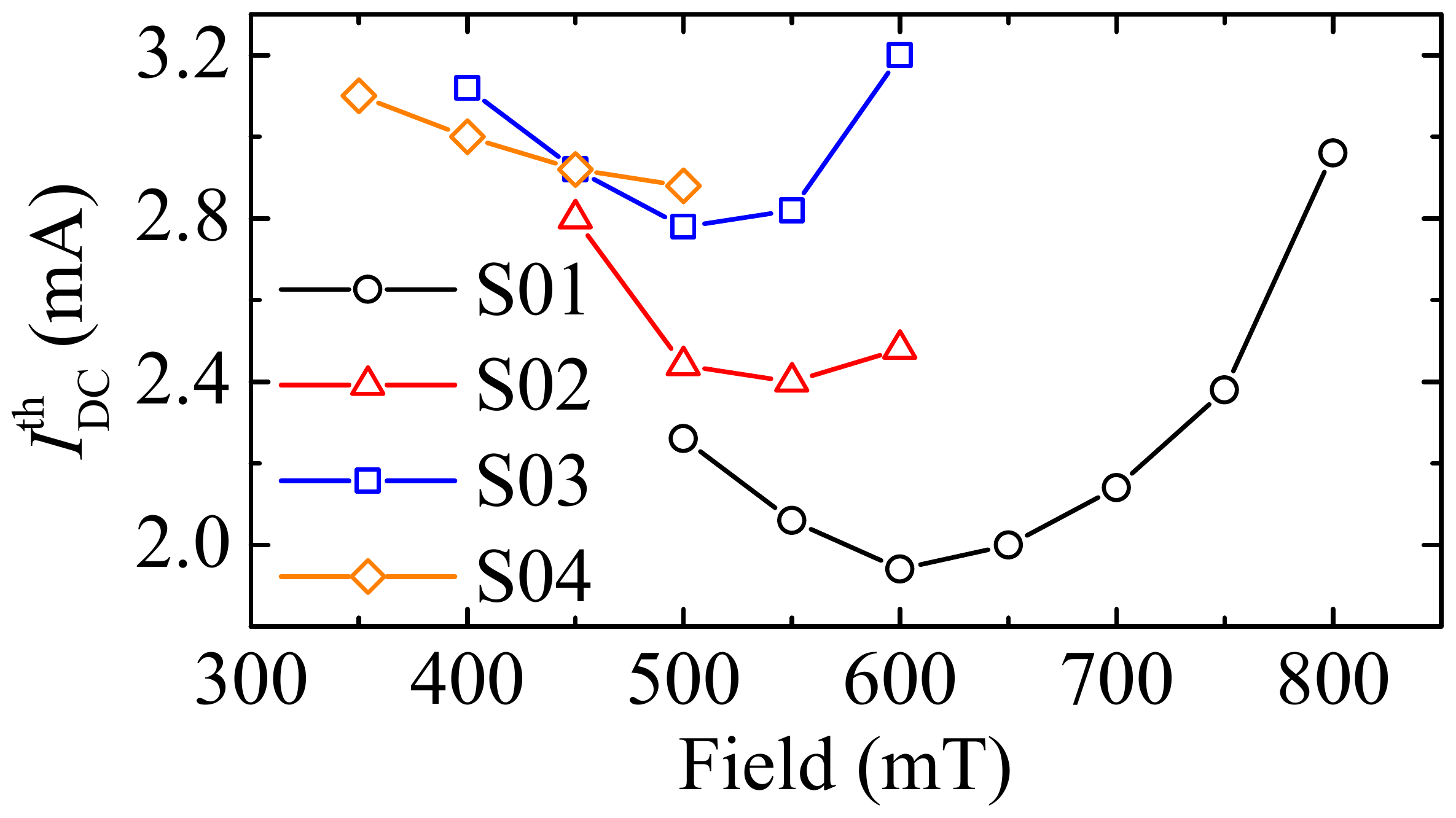}
		\caption{\label{fig:fig5} Threshold current ($I_{\text{DC}}^{\text{th}}$) as a function of external magnetic field for the four devices of this study.}
	\end{figure}
	
	The onset current for auto-oscillations was determined by current scans for external fields ranging between \SI{0.3}{\tesla} and \SI{0.8}{\tesla}, and the results are shown in Fig.~\ref{fig:fig5}. The field dependence of $I_{\text{DC}}^{\text{th}}$ is parabola-like for all samples. This kind of behavior has been predicted in a numerical study by Dvornik \emph{et al.}~\cite{dvornik2018}. The non-monotonic behavior of threshold current as a function of applied field is a result of a re-localization of the auto-oscillation mode and a corresponding change in the spin-transfer-torque (STT) efficiency. In weak oblique magnetic fields, the mode is of edge type and samples a significant portion of the pure spin current, which is largest at the nanoconstriction edges due to the inhomogeneous current density. When the field strength increases, the mode shows an even stronger localization towards the region of the higher current density. Thereby, the STT efficiency increases and the threshold current drops. When the field strength increases further, the mode detaches from the edges and eventually transforms to the bulk type. As this transformation gradually reduces the spatial correlation between the spin current density and the location of the mode, the STT efficiency drops and the threshold current increases. The lower field tunability of $I_{\text{DC}}^{\text{th}}$ of the high damping samples imply an initially larger mode volume, which also was suggested by the evolution of the linewidth.
	
	The field and current range with detectable auto-oscillations is strongly dependent on $\alpha$. The threshold current should increase linearly with damping~\cite{dvornik2018} and the minimum $I_{\text{DC}}^{\text{th}}$  indeed scales with $\alpha$. The enhancement is smaller than predicted (a factor of three), which indicates that the increase in damping is accompanied with a higher STT efficiency. A possible reason for the improved efficiency is a larger SHE through a more transparent interface for alloyed films. The origin of the observed damping dependence of the threshold field is unclear at this stage, calling for a closer inspection of the impact of the applied field on the spectral characteristics.

	\begin{figure}[t]
		\centering
		\includegraphics[width=3.3in]{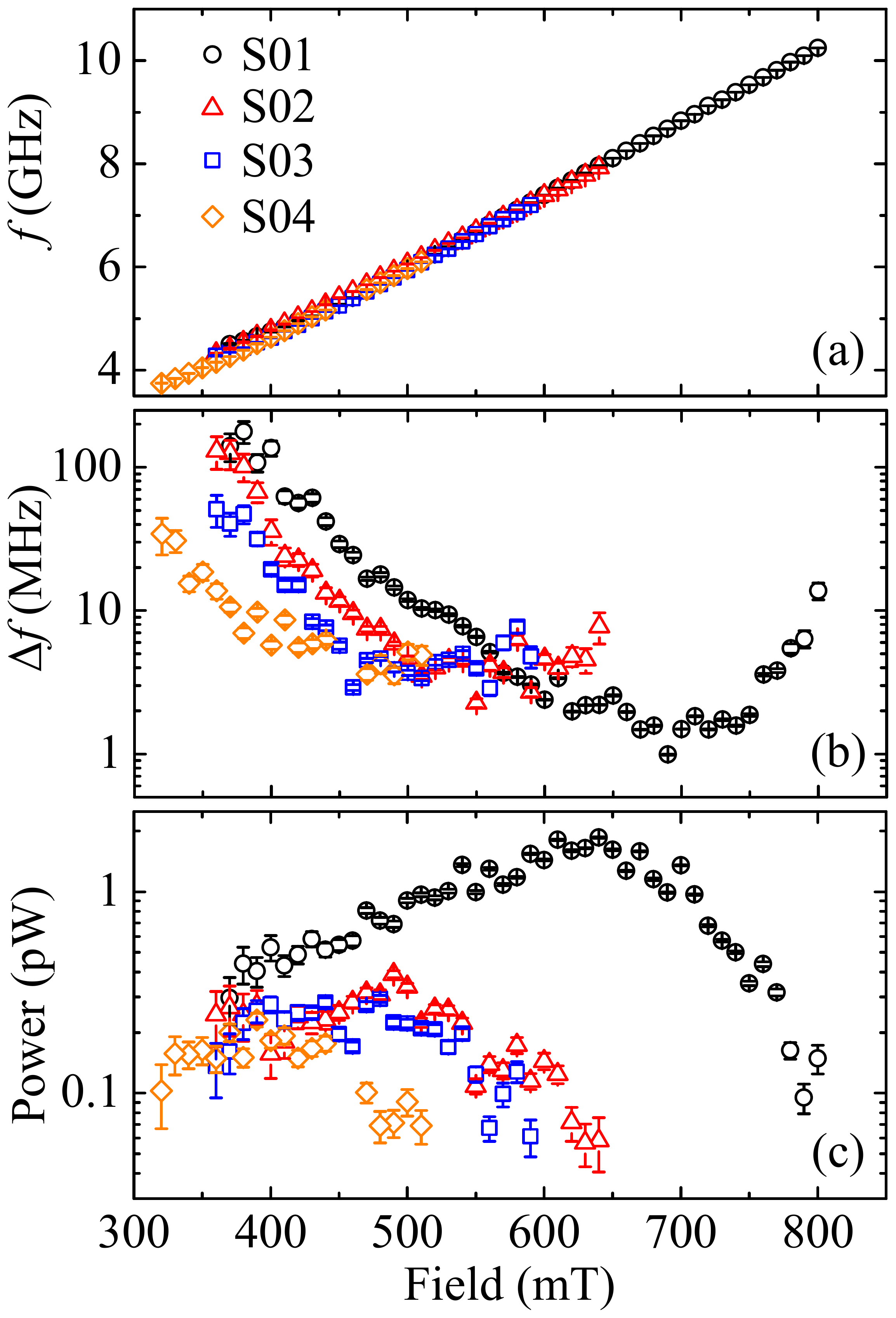}
		\caption{\label{fig:fig6} (a) Frequency, (b) linewidth, and (c) integrated power of the  auto-oscillations as a function of the applied external magnetic field at a constant drive current $I_{\text{DC}} = \SI{3.2}{\mA}$.}
	\end{figure}
	
	Thus, a further investigation of our devices is targeted towards the microwave emission as a function of field with a constant $I_{\text{DC}}$~=~3.2~mA, i.e. above or at the previously measured auto-oscillation threshold for all fields. While the peak frequencies are virtually identical for all the samples, see Fig.~\ref{fig:fig6}(a), the varied damping manifests in a clear pattern in $P$ and $\Delta f$. The microwave power, shown in Fig.~\ref{fig:fig6}(c), firstly increases for all samples with increasing field, peaks for an intermediate field, and finally drops relatively sharp until a point where no more oscillations are detectable. An opposite behavior can be seen for $\Delta f$, which shows a minimum for intermediate fields. The field at which the SHNOs emit their maximum output power decreases monotonically from 0.64~T to 0.4~T with increasing damping. The same trend is visible for the point of minimum linewidth, which decreases with increased damping from 0.71\,T to 0.49\,T, and is therefore at a typically $\sim$0.1~T larger field than the respective maximum power. The lowest overall linewidth can be achieved for the lowest damping SHNO (S01) at high fields, where only this device still shows a detectable signal, i.e., $\Delta f = \SI{1.2}{\MHz}$ at $\mu_{0}H_{\text{ext}} = \SI{0.71}{\tesla}$. However, at low applied fields $\mu_{0}H_{\text{ext}} \leq \SI{0.48}{\tesla}$ a clear trend is noticeable towards smaller linewidths for the alloyed permalloy films with larger damping.
	
	In light of this inverse trend, we can argue that auto-oscillations in nanoconstriction-SHNO should also be described in the framework of \textit{non-linear} auto-oscillators, although the study in Ref.~\onlinecite{dvornik2018} has shown that oscillations in nanoconstriction-SHNOs emerge from a localized \textit{linear} mode. The generation linewidth of nanocontact spin torque oscillators, which are a prime example of non-linear auto-oscillators, has been studied analytically~\cite{kim2008prl1,slavin2009ieeem} and experimentally~\cite{bonetti2012prb}. The linewidth as a function of current and magnetic field angle was shown to follow the expression:
	
	\begin{equation}
	\Delta f = \frac{\Gamma_0}{2 \pi} \left( \frac{k_B T}{E_0} \right) \left[ 1 + \left( \frac{N}{\Gamma_{\text{eff}}} \right)^2 \right] ,
	\label{eq:eq1}
	\end{equation}
	
	where $k_B$, $T$, and $E_0(I_{\text{DC}}/I_{\text{DC}}^{\text{th}})$ are the Boltzmann constant, temperature and the average oscillator energy, respectively. $N$ is the nonlinear frequency shift, a material property that is determined by the internal magnetic field and the magnetization~\cite{gerhart2007prb}. $\Gamma_{\text{eff}}$ is the effective nonlinear damping rate and $\Gamma_0$ is the positive damping rate, and both have an explicit linear dependence on the Gilbert damping $\alpha$~\cite{slavin2009ieeem}. Assuming everything else equal amongst our devices, a decrease of the linewidth with $\alpha$ can be thus expected, when the second term in the brackets in Eq.~\ref{eq:eq1} dominates. This is likely for low to intermediate fields, since $N$ can be calculated to take up the largest values under these conditions~\cite{gerhart2007prb}, which are thus in accordance with the range of fields, where we observe the discussed linewidth vs. damping behavior in our devices.
	
	\section{\label{sec:level6}Conclusions}
	In conclusion, we have fabricated a series of samples where the magnetization is constant, while the spin wave damping is varied by a factor of three. We have shown that the damping of the magnetic layer in nanoconstriction-SHNOs has an important influence on all the spectral characteristics of the devices. The results of our study will encourage the application of tailored materials in SHNOs and can be used for a further understanding of the magnetodynamics in nanodevices, e.g. the coupling mechanisms in mutually synchronized SHNOs.

	\section{Acknowledgments}
	We acknowledge financial support from the China Scholarship Council (CSC), the G\"oran Gustafsson Foundation, the Swedish Research Council (VR), the Knut and Alice Wallenberg Foundation (KAW), and the Swedish Foundation for Strategic Research (SSF). This work was also supported by the European Research Council (ERC) under the European Community’s Seventh Framework Programme (FP/2007-2013)/ERC Grant 307144 ``MUSTANG". 
	

\begin{thebibliography}{28}%
		\makeatletter
		\providecommand \@ifxundefined [1]{%
			\@ifx{#1\undefined}
		}%
		\providecommand \@ifnum [1]{%
			\ifnum #1\expandafter \@firstoftwo
			\else \expandafter \@secondoftwo
			\fi
		}%
		\providecommand \@ifx [1]{%
			\ifx #1\expandafter \@firstoftwo
			\else \expandafter \@secondoftwo
			\fi
		}%
		\providecommand \natexlab [1]{#1}%
		\providecommand \enquote  [1]{``#1''}%
		\providecommand \bibnamefont  [1]{#1}%
		\providecommand \bibfnamefont [1]{#1}%
		\providecommand \citenamefont [1]{#1}%
		\providecommand \href@noop [0]{\@secondoftwo}%
		\providecommand \href [0]{\begingroup \@sanitize@url \@href}%
		\providecommand \@href[1]{\@@startlink{#1}\@@href}%
		\providecommand \@@href[1]{\endgroup#1\@@endlink}%
		\providecommand \@sanitize@url [0]{\catcode `\\12\catcode `\$12\catcode
			`\&12\catcode `\#12\catcode `\^12\catcode `\_12\catcode `\%12\relax}%
		\providecommand \@@startlink[1]{}%
		\providecommand \@@endlink[0]{}%
		\providecommand \url  [0]{\begingroup\@sanitize@url \@url }%
		\providecommand \@url [1]{\endgroup\@href {#1}{\urlprefix }}%
		\providecommand \urlprefix  [0]{URL }%
		\providecommand \Eprint [0]{\href }%
		\providecommand \doibase [0]{http://dx.doi.org/}%
		\providecommand \selectlanguage [0]{\@gobble}%
		\providecommand \bibinfo  [0]{\@secondoftwo}%
		\providecommand \bibfield  [0]{\@secondoftwo}%
		\providecommand \translation [1]{[#1]}%
		\providecommand \BibitemOpen [0]{}%
		\providecommand \bibitemStop [0]{}%
		\providecommand \bibitemNoStop [0]{.\EOS\space}%
		\providecommand \EOS [0]{\spacefactor3000\relax}%
		\providecommand \BibitemShut  [1]{\csname bibitem#1\endcsname}%
		\let\auto@bib@innerbib\@empty
		\bibitem [{\citenamefont {Chen}\ \emph {et~al.}(2016)\citenamefont {Chen},
			\citenamefont {Dumas}, \citenamefont {Eklund}, \citenamefont {Muduli},
			\citenamefont {Houshang}, \citenamefont {Awad}, \citenamefont
			{D\"{u}rrenfeld}, \citenamefont {Malm}, \citenamefont {Rusu},\ and\
			\citenamefont {{\AA}kerman}}]{Chen2016procieee}%
		\BibitemOpen
		\bibfield  {author} {\bibinfo {author} {\bibfnamefont {T.}~\bibnamefont
				{Chen}}, \bibinfo {author} {\bibfnamefont {R.~K.}\ \bibnamefont {Dumas}},
			\bibinfo {author} {\bibfnamefont {A.}~\bibnamefont {Eklund}}, \bibinfo
			{author} {\bibfnamefont {P.~K.}\ \bibnamefont {Muduli}}, \bibinfo {author}
			{\bibfnamefont {A.}~\bibnamefont {Houshang}}, \bibinfo {author}
			{\bibfnamefont {A.~A.}\ \bibnamefont {Awad}}, \bibinfo {author}
			{\bibfnamefont {P.}~\bibnamefont {D\"{u}rrenfeld}}, \bibinfo {author}
			{\bibfnamefont {B.~G.}\ \bibnamefont {Malm}}, \bibinfo {author}
			{\bibfnamefont {A.}~\bibnamefont {Rusu}}, \ and\ \bibinfo {author}
			{\bibfnamefont {J.}~\bibnamefont {{\AA}kerman}},\ }\bibfield  {title}
		{\enquote {\bibinfo {title} {{Spin-Torque and Spin-Hall Nano-Oscillators}},}\
		}\href {\doibase 10.1109/JPROC.2016.2554518} {\bibfield  {journal} {\bibinfo
				{journal} {Proc. IEEE}\ }\textbf {\bibinfo {volume} {104}},\ \bibinfo {pages}
			{1919} (\bibinfo {year} {2016})}\BibitemShut {NoStop}%
		\bibitem [{\citenamefont {{L}iu}\ \emph {et~al.}(2012)\citenamefont {{L}iu},
			\citenamefont {{P}ai}, \citenamefont {{L}i}, \citenamefont {{T}seng},
			\citenamefont {{R}alph},\ and\ \citenamefont {{B}uhrman}}]{liu2012sc}%
		\BibitemOpen
		\bibfield  {author} {\bibinfo {author} {\bibfnamefont {{L}.}~\bibnamefont
				{{L}iu}}, \bibinfo {author} {\bibfnamefont {{C}.-{F}.}\ \bibnamefont
				{{P}ai}}, \bibinfo {author} {\bibfnamefont {{Y}.}~\bibnamefont {{L}i}},
			\bibinfo {author} {\bibfnamefont {{H}.~{W}.}\ \bibnamefont {{T}seng}},
			\bibinfo {author} {\bibfnamefont {{D}.~{C}.}\ \bibnamefont {{R}alph}}, \ and\
			\bibinfo {author} {\bibfnamefont {{R}.~{A}.}\ \bibnamefont {{B}uhrman}},\
		}\bibfield  {title} {\enquote {\bibinfo {title} {{S}pin-{T}orque {S}witching
					with the {G}iant {S}pin {H}all {E}ffect of {T}antalum},}\ }\href {\doibase
			10.1126/science.1218197} {\bibfield  {journal} {\bibinfo  {journal}
				{Science}\ }\textbf {\bibinfo {volume} {336}},\ \bibinfo {pages} {555}
			(\bibinfo {year} {2012})}\BibitemShut {NoStop}%
		\bibitem [{\citenamefont {{D}emidov}\ \emph {et~al.}(2012)\citenamefont
			{{D}emidov}, \citenamefont {{U}razhdin}, \citenamefont {{U}lrichs},
			\citenamefont {{T}iberkevich}, \citenamefont {{S}lavin}, \citenamefont
			{{B}aither}, \citenamefont {{S}chmitz},\ and\ \citenamefont
			{{D}emokritov}}]{demidov2012ntm}%
		\BibitemOpen
		\bibfield  {author} {\bibinfo {author} {\bibfnamefont {{V}.~{E}.}\
				\bibnamefont {{D}emidov}}, \bibinfo {author} {\bibfnamefont
				{{S}.}~\bibnamefont {{U}razhdin}}, \bibinfo {author} {\bibfnamefont
				{{H}.}~\bibnamefont {{U}lrichs}}, \bibinfo {author} {\bibfnamefont
				{{V}.}~\bibnamefont {{T}iberkevich}}, \bibinfo {author} {\bibfnamefont
				{{A}.}~\bibnamefont {{S}lavin}}, \bibinfo {author} {\bibfnamefont
				{{D}.}~\bibnamefont {{B}aither}}, \bibinfo {author} {\bibfnamefont
				{{G}.}~\bibnamefont {{S}chmitz}}, \ and\ \bibinfo {author} {\bibfnamefont
				{{S}.~{O}.}\ \bibnamefont {{D}emokritov}},\ }\bibfield  {title} {\enquote
			{\bibinfo {title} {{M}agnetic nano-oscillator driven by pure spin current},}\
		}\href {http://dx.doi.org/10.1038/nmat3459} {\bibfield  {journal} {\bibinfo
				{journal} {Nat. Mater.}\ }\textbf {\bibinfo {volume} {11}},\ \bibinfo {pages}
			{1028} (\bibinfo {year} {2012})}\BibitemShut {NoStop}%
		\bibitem [{\citenamefont {Demidov}\ \emph {et~al.}(2017)\citenamefont
			{Demidov}, \citenamefont {Urazhdin}, \citenamefont {de~Loubens},
			\citenamefont {Klein}, \citenamefont {Cros}, \citenamefont {Anane},\ and\
			\citenamefont {Demokritov}}]{demidov2017pr}%
		\BibitemOpen
		\bibfield  {author} {\bibinfo {author} {\bibfnamefont {V.E.}\ \bibnamefont
				{Demidov}}, \bibinfo {author} {\bibfnamefont {S.}~\bibnamefont {Urazhdin}},
			\bibinfo {author} {\bibfnamefont {G.}~\bibnamefont {de~Loubens}}, \bibinfo
			{author} {\bibfnamefont {O.}~\bibnamefont {Klein}}, \bibinfo {author}
			{\bibfnamefont {V.}~\bibnamefont {Cros}}, \bibinfo {author} {\bibfnamefont
				{A.}~\bibnamefont {Anane}}, \ and\ \bibinfo {author} {\bibfnamefont {S.O.}\
				\bibnamefont {Demokritov}},\ }\bibfield  {title} {\enquote {\bibinfo {title}
				{Magnetization oscillations and waves driven by pure spin currents},}\ }\href
		{\doibase 10.1016/j.physrep.2017.01.001} {\bibfield  {journal} {\bibinfo
				{journal} {Phys.\ Rev.}\ }\textbf {\bibinfo {volume} {673}},\ \bibinfo
			{pages} {1} (\bibinfo {year} {2017})}\BibitemShut {NoStop}%
		\bibitem [{\citenamefont {{D}emidov}\ \emph {et~al.}(2015)\citenamefont
			{{D}emidov}, \citenamefont {{U}razhdin}, \citenamefont {{D}ivinskiy},
			\citenamefont {{R}inkevich},\ and\ \citenamefont
			{{D}emokritov}}]{demidov2015apl}%
		\BibitemOpen
		\bibfield  {author} {\bibinfo {author} {\bibfnamefont {{V}.~{E}.}\
				\bibnamefont {{D}emidov}}, \bibinfo {author} {\bibfnamefont
				{{S}.}~\bibnamefont {{U}razhdin}}, \bibinfo {author} {\bibfnamefont
				{{B}.}~\bibnamefont {{D}ivinskiy}}, \bibinfo {author} {\bibfnamefont
				{{A}.~{B}.}\ \bibnamefont {{R}inkevich}}, \ and\ \bibinfo {author}
			{\bibfnamefont {{S}.~{O}.}\ \bibnamefont {{D}emokritov}},\ }\bibfield
		{title} {\enquote {\bibinfo {title} {{S}pectral linewidth of spin-current
					nano-oscillators driven by nonlocal spin injection},}\ }\href {\doibase
			http://dx.doi.org/10.1063/1.4936153} {\bibfield  {journal} {\bibinfo
				{journal} {Appl.\ Phys.\ Lett.}\ }\textbf {\bibinfo {volume} {107}},\
			\bibinfo {eid} {202402} (\bibinfo {year} {2015})}\BibitemShut {NoStop}%
		\bibitem [{\citenamefont {Haidar}\ \emph {et~al.}(2016)\citenamefont {Haidar},
			\citenamefont {D{\"{u}}rrenfeld}, \citenamefont {Ranjbar}, \citenamefont
			{Balinsky}, \citenamefont {Fazlali}, \citenamefont {Dvornik}, \citenamefont
			{Dumas}, \citenamefont {Khartsev},\ and\ \citenamefont
			{{\AA}kerman}}]{Haidar2016prb}%
		\BibitemOpen
		\bibfield  {author} {\bibinfo {author} {\bibfnamefont {M.}~\bibnamefont
				{Haidar}}, \bibinfo {author} {\bibfnamefont {P.}~\bibnamefont
				{D{\"{u}}rrenfeld}}, \bibinfo {author} {\bibfnamefont {M.}~\bibnamefont
				{Ranjbar}}, \bibinfo {author} {\bibfnamefont {M.}~\bibnamefont {Balinsky}},
			\bibinfo {author} {\bibfnamefont {M.}~\bibnamefont {Fazlali}}, \bibinfo
			{author} {\bibfnamefont {M.}~\bibnamefont {Dvornik}}, \bibinfo {author}
			{\bibfnamefont {R.~K.}\ \bibnamefont {Dumas}}, \bibinfo {author}
			{\bibfnamefont {S.}~\bibnamefont {Khartsev}}, \ and\ \bibinfo {author}
			{\bibfnamefont {J.}~\bibnamefont {{\AA}kerman}},\ }\bibfield  {title}
		{\enquote {\bibinfo {title} {{Controlling Gilbert damping in a YIG film using
						nonlocal spin currents}},}\ }\href {\doibase 10.1103/PhysRevB.94.180409}
		{\bibfield  {journal} {\bibinfo  {journal} {Phys.\ Rev.\ B}\ }\textbf
			{\bibinfo {volume} {94}},\ \bibinfo {pages} {180409} (\bibinfo {year}
			{2016})}\BibitemShut {NoStop}%
		\bibitem [{\citenamefont {{L}iu}\ \emph {et~al.}(2013)\citenamefont {{L}iu},
			\citenamefont {{L}im},\ and\ \citenamefont {{U}razhdin}}]{liu2013prl}%
		\BibitemOpen
		\bibfield  {author} {\bibinfo {author} {\bibfnamefont {{R}.~{H}.}\
				\bibnamefont {{L}iu}}, \bibinfo {author} {\bibfnamefont {{W}.~{L}.}\
				\bibnamefont {{L}im}}, \ and\ \bibinfo {author} {\bibfnamefont
				{{S}.}~\bibnamefont {{U}razhdin}},\ }\bibfield  {title} {\enquote {\bibinfo
				{title} {{S}pectral {C}haracteristics of the {M}icrowave {E}mission by the
					{S}pin {H}all {N}ano-{O}scillator},}\ }\href {\doibase
			10.1103/PhysRevLett.110.147601} {\bibfield  {journal} {\bibinfo  {journal}
				{Phys.\ Rev.\ Lett.}\ }\textbf {\bibinfo {volume} {110}},\ \bibinfo {pages}
			{147601} (\bibinfo {year} {2013})}\BibitemShut {NoStop}%
		\bibitem [{\citenamefont {{R}anjbar}\ \emph {et~al.}(2014)\citenamefont
			{{R}anjbar}, \citenamefont {{D}\"urrenfeld}, \citenamefont {{H}aidar},
			\citenamefont {{I}acocca}, \citenamefont {{B}alinskiy}, \citenamefont {{L}e},
			\citenamefont {{F}azlali}, \citenamefont {{H}oushang}, \citenamefont
			{{A}wad}, \citenamefont {{D}umas},\ and\ \citenamefont
			{{\AA}kerman}}]{ranjbar2014ieeeml}%
		\BibitemOpen
		\bibfield  {author} {\bibinfo {author} {\bibfnamefont {{M}.}~\bibnamefont
				{{R}anjbar}}, \bibinfo {author} {\bibfnamefont {{P}.}~\bibnamefont
				{{D}\"urrenfeld}}, \bibinfo {author} {\bibfnamefont {{M}.}~\bibnamefont
				{{H}aidar}}, \bibinfo {author} {\bibfnamefont {{E}.}~\bibnamefont
				{{I}acocca}}, \bibinfo {author} {\bibfnamefont {{M}.}~\bibnamefont
				{{B}alinskiy}}, \bibinfo {author} {\bibfnamefont {{T}.{Q}.}\ \bibnamefont
				{{L}e}}, \bibinfo {author} {\bibfnamefont {{M}.}~\bibnamefont {{F}azlali}},
			\bibinfo {author} {\bibfnamefont {{A}.}~\bibnamefont {{H}oushang}}, \bibinfo
			{author} {\bibfnamefont {{A}.{A}.}\ \bibnamefont {{A}wad}}, \bibinfo {author}
			{\bibfnamefont {{R}.{K}.}\ \bibnamefont {{D}umas}}, \ and\ \bibinfo {author}
			{\bibfnamefont {{J}.}~\bibnamefont {{\AA}kerman}},\ }\bibfield  {title}
		{\enquote {\bibinfo {title} {{C}o{F}e{B}-{B}ased {S}pin {H}all
					{N}ano-{O}scillators},}\ }\href {\doibase 10.1109/LMAG.2014.2375155}
		{\bibfield  {journal} {\bibinfo  {journal} {IEEE Magn. Lett.}\ }\textbf
			{\bibinfo {volume} {5}},\ \bibinfo {pages} {3000504} (\bibinfo {year}
			{2014})}\BibitemShut {NoStop}%
		\bibitem [{\citenamefont {Divinskiy}\ \emph {et~al.}(2017)\citenamefont
			{Divinskiy}, \citenamefont {Demidov}, \citenamefont {Kozhanov}, \citenamefont
			{Rinkevich}, \citenamefont {Demokritov},\ and\ \citenamefont
			{Urazhdin}}]{divinskiy2017apl}%
		\BibitemOpen
		\bibfield  {author} {\bibinfo {author} {\bibfnamefont {B.}~\bibnamefont
				{Divinskiy}}, \bibinfo {author} {\bibfnamefont {V.~E.}\ \bibnamefont
				{Demidov}}, \bibinfo {author} {\bibfnamefont {A.}~\bibnamefont {Kozhanov}},
			\bibinfo {author} {\bibfnamefont {A.~B.}\ \bibnamefont {Rinkevich}}, \bibinfo
			{author} {\bibfnamefont {S.~O.}\ \bibnamefont {Demokritov}}, \ and\ \bibinfo
			{author} {\bibfnamefont {S.}~\bibnamefont {Urazhdin}},\ }\bibfield  {title}
		{\enquote {\bibinfo {title} {Nanoconstriction spin-hall oscillator with
					perpendicular magnetic anisotropy},}\ }\href {\doibase 10.1063/1.4993910}
		{\bibfield  {journal} {\bibinfo  {journal} {Appl.\ Phys.\ Lett.}\ }\textbf
			{\bibinfo {volume} {111}},\ \bibinfo {pages} {032405} (\bibinfo {year}
			{2017})}\BibitemShut {NoStop}%
		\bibitem [{\citenamefont {{D}emidov}\ \emph {et~al.}(2014)\citenamefont
			{{D}emidov}, \citenamefont {{U}razhdin}, \citenamefont {{Z}holud},
			\citenamefont {{S}adovnikov},\ and\ \citenamefont
			{{D}emokritov}}]{demidov2014apl}%
		\BibitemOpen
		\bibfield  {author} {\bibinfo {author} {\bibfnamefont {{V}.~{E}.}\
				\bibnamefont {{D}emidov}}, \bibinfo {author} {\bibfnamefont
				{{S}.}~\bibnamefont {{U}razhdin}}, \bibinfo {author} {\bibfnamefont
				{{A}.}~\bibnamefont {{Z}holud}}, \bibinfo {author} {\bibfnamefont
				{{A}.~{V}.}\ \bibnamefont {{S}adovnikov}}, \ and\ \bibinfo {author}
			{\bibfnamefont {{S}.~{O}.}\ \bibnamefont {{D}emokritov}},\ }\bibfield
		{title} {\enquote {\bibinfo {title} {{N}anoconstriction-based spin-{H}all
					nano-oscillator},}\ }\href {\doibase http://dx.doi.org/10.1063/1.4901027}
		{\bibfield  {journal} {\bibinfo  {journal} {Appl.\ Phys.\ Lett.}\ }\textbf
			{\bibinfo {volume} {105}},\ \bibinfo {eid} {172410} (\bibinfo {year}
			{2014})}\BibitemShut {NoStop}%
		\bibitem [{\citenamefont {D\"{u}rrenfeld}\ \emph {et~al.}(2017)\citenamefont
			{D\"{u}rrenfeld}, \citenamefont {Awad}, \citenamefont {Houshang},
			\citenamefont {Dumas},\ and\ \citenamefont
			{{\AA}kerman}}]{duerrenfeld2017ns}%
		\BibitemOpen
		\bibfield  {author} {\bibinfo {author} {\bibfnamefont {P.}~\bibnamefont
				{D\"{u}rrenfeld}}, \bibinfo {author} {\bibfnamefont {A.~A.}\ \bibnamefont
				{Awad}}, \bibinfo {author} {\bibfnamefont {A.}~\bibnamefont {Houshang}},
			\bibinfo {author} {\bibfnamefont {R.~K.}\ \bibnamefont {Dumas}}, \ and\
			\bibinfo {author} {\bibfnamefont {J.}~\bibnamefont {{\AA}kerman}},\
		}\bibfield  {title} {\enquote {\bibinfo {title} {{A} 20 nm spin {H}all
					nano-oscillator},}\ }\href {\doibase 10.1039/C6NR07903B} {\bibfield
			{journal} {\bibinfo  {journal} {Nanoscale}\ }\textbf {\bibinfo {volume}
				{9}},\ \bibinfo {pages} {1285} (\bibinfo {year} {2017})}\BibitemShut
		{NoStop}%
		\bibitem [{\citenamefont {Mazraati}\ \emph {et~al.}(2016)\citenamefont
			{Mazraati}, \citenamefont {Chung}, \citenamefont {Houshang}, \citenamefont
			{Dvornik}, \citenamefont {Piazza}, \citenamefont {Qejvanaj}, \citenamefont
			{Jiang}, \citenamefont {Le}, \citenamefont {Weissenrieder},\ and\
			\citenamefont {{\AA}kerman}}]{mazraati2016apl}%
		\BibitemOpen
		\bibfield  {author} {\bibinfo {author} {\bibfnamefont {H.}~\bibnamefont
				{Mazraati}}, \bibinfo {author} {\bibfnamefont {S.}~\bibnamefont {Chung}},
			\bibinfo {author} {\bibfnamefont {A.}~\bibnamefont {Houshang}}, \bibinfo
			{author} {\bibfnamefont {M.}~\bibnamefont {Dvornik}}, \bibinfo {author}
			{\bibfnamefont {L.}~\bibnamefont {Piazza}}, \bibinfo {author} {\bibfnamefont
				{F.}~\bibnamefont {Qejvanaj}}, \bibinfo {author} {\bibfnamefont
				{S.}~\bibnamefont {Jiang}}, \bibinfo {author} {\bibfnamefont {T.~Q.}\
				\bibnamefont {Le}}, \bibinfo {author} {\bibfnamefont {J.}~\bibnamefont
				{Weissenrieder}}, \ and\ \bibinfo {author} {\bibfnamefont {J.}~\bibnamefont
				{{\AA}kerman}},\ }\bibfield  {title} {\enquote {\bibinfo {title} {{L}ow
					operational current spin {H}all nano-oscillators based on {NiFe/W}
					bilayers},}\ }\href {\doibase 10.1063/1.4971828} {\bibfield  {journal}
			{\bibinfo  {journal} {Appl.\ Phys.\ Lett.}\ }\textbf {\bibinfo {volume}
				{109}},\ \bibinfo {pages} {242402} (\bibinfo {year} {2016})}\BibitemShut
		{NoStop}%
		\bibitem [{\citenamefont {Dvornik}\ \emph {et~al.}(2018)\citenamefont
			{Dvornik}, \citenamefont {Awad},\ and\ \citenamefont
			{{\AA}kerman}}]{dvornik2018}%
		\BibitemOpen
		\bibfield  {author} {\bibinfo {author} {\bibfnamefont {M.}~\bibnamefont
				{Dvornik}}, \bibinfo {author} {\bibfnamefont {A.~A.}\ \bibnamefont {Awad}}, \
			and\ \bibinfo {author} {\bibfnamefont {J.}~\bibnamefont {{\AA}kerman}},\
		}\bibfield  {title} {\enquote {\bibinfo {title} {Origin of magnetization
					auto-oscillations in constriction-based spin hall nano-oscillators},}\ }\href
		{\doibase 10.1103/PhysRevApplied.9.014017} {\bibfield  {journal} {\bibinfo
				{journal} {Phys. Rev. Appl.}\ }\textbf {\bibinfo {volume} {9}},\ \bibinfo
			{pages} {014017} (\bibinfo {year} {2018})}\BibitemShut {NoStop}%
		\bibitem [{\citenamefont {{D}uan}\ \emph {et~al.}(2014)\citenamefont {{D}uan},
			\citenamefont {{S}mith}, \citenamefont {{Y}ang}, \citenamefont
			{{Y}oungblood}, \citenamefont {{L}indner}, \citenamefont {{D}emidov},
			\citenamefont {{D}emokritov},\ and\ \citenamefont
			{{K}rivorotov}}]{duan2014ntc}%
		\BibitemOpen
		\bibfield  {author} {\bibinfo {author} {\bibfnamefont {{Z}.}~\bibnamefont
				{{D}uan}}, \bibinfo {author} {\bibfnamefont {{A}.}~\bibnamefont {{S}mith}},
			\bibinfo {author} {\bibfnamefont {{L}.}~\bibnamefont {{Y}ang}}, \bibinfo
			{author} {\bibfnamefont {{B}.}~\bibnamefont {{Y}oungblood}}, \bibinfo
			{author} {\bibfnamefont {{J}.}~\bibnamefont {{L}indner}}, \bibinfo {author}
			{\bibfnamefont {{V}.~{E}.}\ \bibnamefont {{D}emidov}}, \bibinfo {author}
			{\bibfnamefont {{S}.~{O}.}\ \bibnamefont {{D}emokritov}}, \ and\ \bibinfo
			{author} {\bibfnamefont {{I}.~{N}.}\ \bibnamefont {{K}rivorotov}},\
		}\bibfield  {title} {\enquote {\bibinfo {title} {{N}anowire spin torque
					oscillator driven by spin orbit torques},}\ }\href
		{http://dx.doi.org/10.1038/ncomms6616} {\bibfield  {journal} {\bibinfo
				{journal} {Nat.\ Commun.}\ }\textbf {\bibinfo {volume} {5}},\ \bibinfo
			{pages} {5616} (\bibinfo {year} {2014})}\BibitemShut {NoStop}%
		\bibitem [{\citenamefont {{Y}ang}\ \emph {et~al.}(2015)\citenamefont {{Y}ang},
			\citenamefont {{V}erba}, \citenamefont {{T}iberkevich}, \citenamefont
			{{S}chneider}, \citenamefont {{S}mith}, \citenamefont {{D}uan}, \citenamefont
			{{Y}oungblood}, \citenamefont {{L}enz}, \citenamefont {{L}indner},
			\citenamefont {{S}lavin},\ and\ \citenamefont {{K}rivorotov}}]{yang2015srp}%
		\BibitemOpen
		\bibfield  {author} {\bibinfo {author} {\bibfnamefont {{L}.}~\bibnamefont
				{{Y}ang}}, \bibinfo {author} {\bibfnamefont {{R}.}~\bibnamefont {{V}erba}},
			\bibinfo {author} {\bibfnamefont {{V}.}~\bibnamefont {{T}iberkevich}},
			\bibinfo {author} {\bibfnamefont {{T}.}~\bibnamefont {{S}chneider}}, \bibinfo
			{author} {\bibfnamefont {{A}.}~\bibnamefont {{S}mith}}, \bibinfo {author}
			{\bibfnamefont {{Z}.}~\bibnamefont {{D}uan}}, \bibinfo {author}
			{\bibfnamefont {{B}.}~\bibnamefont {{Y}oungblood}}, \bibinfo {author}
			{\bibfnamefont {{K}.}~\bibnamefont {{L}enz}}, \bibinfo {author}
			{\bibfnamefont {{J}.}~\bibnamefont {{L}indner}}, \bibinfo {author}
			{\bibfnamefont {{A}.~{N}.}\ \bibnamefont {{S}lavin}}, \ and\ \bibinfo
			{author} {\bibfnamefont {{I}.~{N}.}\ \bibnamefont {{K}rivorotov}},\
		}\bibfield  {title} {\enquote {\bibinfo {title} {{R}eduction of phase noise
					in nanowire spin orbit torque oscillators},}\ }\href
		{http://dx.doi.org/10.1038/srep16942} {\bibfield  {journal} {\bibinfo
				{journal} {Sci. Rep.}\ }\textbf {\bibinfo {volume} {5}},\ \bibinfo {pages}
			{16942} (\bibinfo {year} {2015})}\BibitemShut {NoStop}%
		\bibitem [{\citenamefont {{K}endziorczyk}\ and\ \citenamefont
			{{K}uhn}(2016)}]{kendziorczyk2016prb}%
		\BibitemOpen
		\bibfield  {author} {\bibinfo {author} {\bibfnamefont {{T}.}~\bibnamefont
				{{K}endziorczyk}}\ and\ \bibinfo {author} {\bibfnamefont {{T}.}~\bibnamefont
				{{K}uhn}},\ }\bibfield  {title} {\enquote {\bibinfo {title} {{M}utual
					synchronization of nanoconstriction-based spin {H}all nano-oscillators
					through evanescent and propagating spin waves},}\ }\href {\doibase
			10.1103/PhysRevB.93.134413} {\bibfield  {journal} {\bibinfo  {journal}
				{Phys.\ Rev.\ B}\ }\textbf {\bibinfo {volume} {93}},\ \bibinfo {pages}
			{134413} (\bibinfo {year} {2016})}\BibitemShut {NoStop}%
		\bibitem [{\citenamefont {Awad}\ \emph {et~al.}(2017)\citenamefont {Awad},
			\citenamefont {D\"{u}rrenfeld}, \citenamefont {Houshang}, \citenamefont
			{Dvornik}, \citenamefont {Iacocca}, \citenamefont {Dumas},\ and\
			\citenamefont {{\AA}kerman}}]{awad2017ntp}%
		\BibitemOpen
		\bibfield  {author} {\bibinfo {author} {\bibfnamefont {A.~A.}\ \bibnamefont
				{Awad}}, \bibinfo {author} {\bibfnamefont {P.}~\bibnamefont
				{D\"{u}rrenfeld}}, \bibinfo {author} {\bibfnamefont {A.}~\bibnamefont
				{Houshang}}, \bibinfo {author} {\bibfnamefont {M.}~\bibnamefont {Dvornik}},
			\bibinfo {author} {\bibfnamefont {E.}~\bibnamefont {Iacocca}}, \bibinfo
			{author} {\bibfnamefont {R.~K.}\ \bibnamefont {Dumas}}, \ and\ \bibinfo
			{author} {\bibfnamefont {J.}~\bibnamefont {{\AA}kerman}},\ }\bibfield
		{title} {\enquote {\bibinfo {title} {{L}ong-range mutual synchronization of
					spin {H}all nano-oscillators},}\ }\href {http://dx.doi.org/10.1038/nphys3927}
		{\bibfield  {journal} {\bibinfo  {journal} {Nat.\ Phys.}\ }\textbf {\bibinfo
				{volume} {13}},\ \bibinfo {pages} {292} (\bibinfo {year} {2017})}\BibitemShut
		{NoStop}%
		\bibitem [{\citenamefont {{Y}in}\ \emph {et~al.}(2015)\citenamefont {{Y}in},
			\citenamefont {{P}an}, \citenamefont {{A}hlberg}, \citenamefont {{R}anjbar},
			\citenamefont {{D}\"urrenfeld}, \citenamefont {{H}oushang}, \citenamefont
			{{H}aidar}, \citenamefont {{B}ergqvist}, \citenamefont {{Z}hai},
			\citenamefont {{D}umas}, \citenamefont {{D}elin},\ and\ \citenamefont
			{{\AA}kerman}}]{yin2015prb}%
		\BibitemOpen
		\bibfield  {author} {\bibinfo {author} {\bibfnamefont {{Y}.}~\bibnamefont
				{{Y}in}}, \bibinfo {author} {\bibfnamefont {{F}.}~\bibnamefont {{P}an}},
			\bibinfo {author} {\bibfnamefont {{M}.}~\bibnamefont {{A}hlberg}}, \bibinfo
			{author} {\bibfnamefont {{M}.}~\bibnamefont {{R}anjbar}}, \bibinfo {author}
			{\bibfnamefont {{P}.}~\bibnamefont {{D}\"urrenfeld}}, \bibinfo {author}
			{\bibfnamefont {{A}.}~\bibnamefont {{H}oushang}}, \bibinfo {author}
			{\bibfnamefont {{M}.}~\bibnamefont {{H}aidar}}, \bibinfo {author}
			{\bibfnamefont {{L}.}~\bibnamefont {{B}ergqvist}}, \bibinfo {author}
			{\bibfnamefont {{Y}.}~\bibnamefont {{Z}hai}}, \bibinfo {author}
			{\bibfnamefont {{R}.~{K}.}\ \bibnamefont {{D}umas}}, \bibinfo {author}
			{\bibfnamefont {{A}.}~\bibnamefont {{D}elin}}, \ and\ \bibinfo {author}
			{\bibfnamefont {{J}.}~\bibnamefont {{\AA}kerman}},\ }\bibfield  {title}
		{\enquote {\bibinfo {title} {{T}unable permalloy-based films for magnonic
					devices},}\ }\href {\doibase 10.1103/PhysRevB.92.024427} {\bibfield
			{journal} {\bibinfo  {journal} {Phys.\ Rev.\ B}\ }\textbf {\bibinfo {volume}
				{92}},\ \bibinfo {pages} {024427} (\bibinfo {year} {2015})}\BibitemShut
		{NoStop}%
		\bibitem [{\citenamefont {{T}serkovnyak}\ \emph {et~al.}(2002)\citenamefont
			{{T}serkovnyak}, \citenamefont {{B}rataas},\ and\ \citenamefont
			{{B}auer}}]{tserkovnyak2002prl}%
		\BibitemOpen
		\bibfield  {author} {\bibinfo {author} {\bibfnamefont {{Y}.}~\bibnamefont
				{{T}serkovnyak}}, \bibinfo {author} {\bibfnamefont {{A}.}~\bibnamefont
				{{B}rataas}}, \ and\ \bibinfo {author} {\bibfnamefont {{G}. {E}.~{W}.}\
				\bibnamefont {{B}auer}},\ }\bibfield  {title} {\enquote {\bibinfo {title}
				{{E}nhanced {G}ilbert {D}amping in {T}hin {F}erromagnetic {F}ilms},}\ }\href
		{\doibase 10.1103/PhysRevLett.88.117601} {\bibfield  {journal} {\bibinfo
				{journal} {Phys.\ Rev.\ Lett.}\ }\textbf {\bibinfo {volume} {88}},\ \bibinfo
			{pages} {117601} (\bibinfo {year} {2002})}\BibitemShut {NoStop}%
		\bibitem [{\citenamefont {{M}izukami}\ \emph {et~al.}(2002)\citenamefont
			{{M}izukami}, \citenamefont {{A}ndo},\ and\ \citenamefont
			{{M}iyazaki}}]{mizukami2002prb}%
		\BibitemOpen
		\bibfield  {author} {\bibinfo {author} {\bibfnamefont {{S}.}~\bibnamefont
				{{M}izukami}}, \bibinfo {author} {\bibfnamefont {{Y}.}~\bibnamefont
				{{A}ndo}}, \ and\ \bibinfo {author} {\bibfnamefont {{T}.}~\bibnamefont
				{{M}iyazaki}},\ }\bibfield  {title} {\enquote {\bibinfo {title} {{E}ffect of
					spin diffusion on {G}ilbert damping for a very thin permalloy layer in
					{C}u/permalloy/{C}u/{P}t films},}\ }\href {\doibase
			10.1103/PhysRevB.66.104413} {\bibfield  {journal} {\bibinfo  {journal}
				{Phys.\ Rev.\ B}\ }\textbf {\bibinfo {volume} {66}},\ \bibinfo {pages}
			{104413} (\bibinfo {year} {2002})}\BibitemShut {NoStop}%
		\bibitem [{\citenamefont {{S}un}\ \emph {et~al.}(2013)\citenamefont {{S}un},
			\citenamefont {{C}hang}, \citenamefont {{K}abatek}, \citenamefont {{S}ong},
			\citenamefont {{W}ang}, \citenamefont {{J}antz}, \citenamefont {{S}chneider},
			\citenamefont {{W}u}, \citenamefont {{M}ontoya}, \citenamefont {{K}ardasz},
			\citenamefont {{H}einrich}, \citenamefont {te~{V}elthuis}, \citenamefont
			{{S}chultheiss},\ and\ \citenamefont {{H}offmann}}]{sun2013prl}%
		\BibitemOpen
		\bibfield  {author} {\bibinfo {author} {\bibfnamefont {{Y}.}~\bibnamefont
				{{S}un}}, \bibinfo {author} {\bibfnamefont {{H}.}~\bibnamefont {{C}hang}},
			\bibinfo {author} {\bibfnamefont {{M}.}~\bibnamefont {{K}abatek}}, \bibinfo
			{author} {\bibfnamefont {{Y}.-{Y}.}\ \bibnamefont {{S}ong}}, \bibinfo
			{author} {\bibfnamefont {{Z}.}~\bibnamefont {{W}ang}}, \bibinfo {author}
			{\bibfnamefont {{M}.}~\bibnamefont {{J}antz}}, \bibinfo {author}
			{\bibfnamefont {{W}.}~\bibnamefont {{S}chneider}}, \bibinfo {author}
			{\bibfnamefont {{M}.}~\bibnamefont {{W}u}}, \bibinfo {author} {\bibfnamefont
				{{E}.}~\bibnamefont {{M}ontoya}}, \bibinfo {author} {\bibfnamefont
				{{B}.}~\bibnamefont {{K}ardasz}}, \bibinfo {author} {\bibfnamefont
				{{B}.}~\bibnamefont {{H}einrich}}, \bibinfo {author} {\bibfnamefont {{S}.
					{G}.~{E}.}\ \bibnamefont {te~{V}elthuis}}, \bibinfo {author} {\bibfnamefont
				{{H}.}~\bibnamefont {{S}chultheiss}}, \ and\ \bibinfo {author} {\bibfnamefont
				{{A}.}~\bibnamefont {{H}offmann}},\ }\bibfield  {title} {\enquote {\bibinfo
				{title} {{D}amping in {Y}ttrium {I}ron {G}arnet {N}anoscale {F}ilms {C}apped
					by {P}latinum},}\ }\href {\doibase 10.1103/PhysRevLett.111.106601} {\bibfield
			{journal} {\bibinfo  {journal} {Phys.\ Rev.\ Lett.}\ }\textbf {\bibinfo
				{volume} {111}},\ \bibinfo {pages} {106601} (\bibinfo {year}
			{2013})}\BibitemShut {NoStop}%
		\bibitem [{\citenamefont {{L}iu}\ \emph {et~al.}(2011)\citenamefont {{L}iu},
			\citenamefont {{M}oriyama}, \citenamefont {{R}alph},\ and\ \citenamefont
			{{B}uhrman}}]{liu2011prl}%
		\BibitemOpen
		\bibfield  {author} {\bibinfo {author} {\bibfnamefont {{L}.}~\bibnamefont
				{{L}iu}}, \bibinfo {author} {\bibfnamefont {{T}.}~\bibnamefont {{M}oriyama}},
			\bibinfo {author} {\bibfnamefont {{D}.~{C}.}\ \bibnamefont {{R}alph}}, \ and\
			\bibinfo {author} {\bibfnamefont {{R}.~{A}.}\ \bibnamefont {{B}uhrman}},\
		}\bibfield  {title} {\enquote {\bibinfo {title} {{S}pin-{T}orque
					{F}erromagnetic {R}esonance {I}nduced by the {S}pin {H}all {E}ffect},}\
		}\href {\doibase 10.1103/PhysRevLett.106.036601} {\bibfield  {journal}
			{\bibinfo  {journal} {Phys.\ Rev.\ Lett.}\ }\textbf {\bibinfo {volume}
				{106}},\ \bibinfo {pages} {036601} (\bibinfo {year} {2011})}\BibitemShut
		{NoStop}%
		\bibitem [{\citenamefont {{S}lavin}\ and\ \citenamefont
			{{T}iberkevich}(2009)}]{slavin2009ieeem}%
		\BibitemOpen
		\bibfield  {author} {\bibinfo {author} {\bibfnamefont {{A}.}~\bibnamefont
				{{S}lavin}}\ and\ \bibinfo {author} {\bibfnamefont {{V}.}~\bibnamefont
				{{T}iberkevich}},\ }\bibfield  {title} {\enquote {\bibinfo {title}
				{{N}onlinear {A}uto-{O}scillator {T}heory of {M}icrowave {G}eneration by
					{S}pin-{P}olarized {C}urrent},}\ }\href {\doibase 10.1109/TMAG.2008.2009935}
		{\bibfield  {journal} {\bibinfo  {journal} {IEEE Trans. Magn.}\ }\textbf
			{\bibinfo {volume} {45}},\ \bibinfo {pages} {1875} (\bibinfo {year}
			{2009})}\BibitemShut {NoStop}%
		\bibitem [{\citenamefont {Tiberkevich}\ \emph {et~al.}(2009)\citenamefont
			{Tiberkevich}, \citenamefont {Slavin}, \citenamefont {Bankowski},\ and\
			\citenamefont {Gerhart}}]{tiberkevich2009apl}%
		\BibitemOpen
		\bibfield  {author} {\bibinfo {author} {\bibfnamefont {V.}~\bibnamefont
				{Tiberkevich}}, \bibinfo {author} {\bibfnamefont {A.}~\bibnamefont {Slavin}},
			\bibinfo {author} {\bibfnamefont {E.}~\bibnamefont {Bankowski}}, \ and\
			\bibinfo {author} {\bibfnamefont {G.}~\bibnamefont {Gerhart}},\ }\bibfield
		{title} {\enquote {\bibinfo {title} {Phase-locking and frustration in an
					array of nonlinear spin-torque nano-oscillators},}\ }\href {\doibase
			10.1063/1.3278602} {\bibfield  {journal} {\bibinfo  {journal} {Appl.\ Phys.\
					Lett.}\ }\textbf {\bibinfo {volume} {95}},\ \bibinfo {pages} {262505}
			(\bibinfo {year} {2009})}\BibitemShut {NoStop}%
		\bibitem [{\citenamefont {Tiberkevich}\ \emph {et~al.}(2014)\citenamefont
			{Tiberkevich}, \citenamefont {Khymyn}, \citenamefont {Tang},\ and\
			\citenamefont {Slavin}}]{tiberkevich2014srp}%
		\BibitemOpen
		\bibfield  {author} {\bibinfo {author} {\bibfnamefont {V.~S.}\ \bibnamefont
				{Tiberkevich}}, \bibinfo {author} {\bibfnamefont {R.~S.}\ \bibnamefont
				{Khymyn}}, \bibinfo {author} {\bibfnamefont {H.~X.}\ \bibnamefont {Tang}}, \
			and\ \bibinfo {author} {\bibfnamefont {A.~N.}\ \bibnamefont {Slavin}},\
		}\bibfield  {title} {\enquote {\bibinfo {title} {Sensitivity to external
					signals and synchronization properties of a non-isochronous auto-oscillator
					with delayed feedback},}\ }\href@noop {} {\bibfield  {journal} {\bibinfo
				{journal} {Sci. Rep.}\ }\textbf {\bibinfo {volume} {4}},\ \bibinfo {pages}
			{3873} (\bibinfo {year} {2014})}\BibitemShut {NoStop}%
		\bibitem [{\citenamefont {{Kim}}\ \emph {et~al.}(2008)\citenamefont {{Kim}},
			\citenamefont {{Tiberkevich}},\ and\ \citenamefont {{Slavin}}}]{kim2008prl1}%
		\BibitemOpen
		\bibfield  {author} {\bibinfo {author} {\bibfnamefont {{J}.-{V}.}\
				\bibnamefont {{Kim}}}, \bibinfo {author} {\bibfnamefont {{V}.}~\bibnamefont
				{{Tiberkevich}}}, \ and\ \bibinfo {author} {\bibfnamefont {{A}.~{N}.}\
				\bibnamefont {{Slavin}}},\ }\bibfield  {title} {\enquote {\bibinfo {title}
				{{Generation Linewidth of an Auto-Oscillator with a Nonlinear Frequency
						Shift: Spin-Torque Nano-Oscillator}},}\ }\href {\doibase
			10.1103/PhysRevLett.100.017207} {\bibfield  {journal} {\bibinfo  {journal}
				{Phys.\ Rev.\ Lett.}\ }\textbf {\bibinfo {volume} {100}},\ \bibinfo {pages}
			{017207} (\bibinfo {year} {2008})}\BibitemShut {NoStop}%
		\bibitem [{\citenamefont {{B}onetti}\ \emph {et~al.}(2012)\citenamefont
			{{B}onetti}, \citenamefont {{P}uliafito}, \citenamefont {{C}onsolo},
			\citenamefont {{T}iberkevich}, \citenamefont {{S}lavin},\ and\ \citenamefont
			{{\AA}kerman}}]{bonetti2012prb}%
		\BibitemOpen
		\bibfield  {author} {\bibinfo {author} {\bibfnamefont {{S}.}~\bibnamefont
				{{B}onetti}}, \bibinfo {author} {\bibfnamefont {{V}.}~\bibnamefont
				{{P}uliafito}}, \bibinfo {author} {\bibfnamefont {{G}.}~\bibnamefont
				{{C}onsolo}}, \bibinfo {author} {\bibfnamefont {{V}.~{S}.}\ \bibnamefont
				{{T}iberkevich}}, \bibinfo {author} {\bibfnamefont {{A}.~{N}.}\ \bibnamefont
				{{S}lavin}}, \ and\ \bibinfo {author} {\bibfnamefont {{J}.}~\bibnamefont
				{{\AA}kerman}},\ }\bibfield  {title} {\enquote {\bibinfo {title} {{P}ower and
					linewidth of propagating and localized modes in nanocontact spin-torque
					oscillators},}\ }\href {\doibase 10.1103/PhysRevB.85.174427} {\bibfield
			{journal} {\bibinfo  {journal} {Phys.\ Rev.\ B}\ }\textbf {\bibinfo {volume}
				{85}},\ \bibinfo {pages} {174427} (\bibinfo {year} {2012})}\BibitemShut
		{NoStop}%
		\bibitem [{\citenamefont {{Gerhart}}\ \emph {et~al.}(2007)\citenamefont
			{{Gerhart}}, \citenamefont {{Bankowski}}, \citenamefont {{Melkov}},
			\citenamefont {{Tiberkevich}},\ and\ \citenamefont
			{{Slavin}}}]{gerhart2007prb}%
		\BibitemOpen
		\bibfield  {author} {\bibinfo {author} {\bibfnamefont {{G}.}~\bibnamefont
				{{Gerhart}}}, \bibinfo {author} {\bibfnamefont {{E}.}~\bibnamefont
				{{Bankowski}}}, \bibinfo {author} {\bibfnamefont {{G}.~{A}.}\ \bibnamefont
				{{Melkov}}}, \bibinfo {author} {\bibfnamefont {{V}.~{S}.}\ \bibnamefont
				{{Tiberkevich}}}, \ and\ \bibinfo {author} {\bibfnamefont {{A}.~{N}.}\
				\bibnamefont {{Slavin}}},\ }\bibfield  {title} {\enquote {\bibinfo {title}
				{{Angular dependence of the microwave-generation threshold in a nanoscale
						spin-torque oscillator}},}\ }\href {\doibase 10.1103/PhysRevB.76.024437}
		{\bibfield  {journal} {\bibinfo  {journal} {Phys.\ Rev.\ B}\ }\textbf
			{\bibinfo {volume} {76}},\ \bibinfo {pages} {024437} (\bibinfo {year}
			{2007})}\BibitemShut {NoStop}%
	\end{thebibliography}
	
	%
	
\end{document}